\documentclass[cameraready]{Interspeech}
\usepackage{amsmath,graphicx,hyperref}
\usepackage{booktabs}
\usepackage{siunitx}
\usepackage{caption}
\usepackage{subcaption}
\usepackage{cite}
\usepackage{comment}
\newcommand{\zvi}[1]{\mathbf{z}_{#1}}

\title{Towards Audio Token Compression in Large Audio Language Models}
\author[affiliation={1},correspondingauthor]{Saurabhchand}{Bhati}
\author[affiliation={2,3}]{Samuel}{Thomas}
\author[affiliation={3,4}]{Hilde}{Kuehne}
\author[affiliation={2,3}]{Rogerio}{Feris}
\author[affiliation={1}]{James}{Glass}
\address{
$^1$ MIT, USA \\
$^{2}$IBM Research\\
$^{3}$MIT-IBM Watson AI Lab \\
$^{4}$Tuebingen AI Center/University of Tuebingen}
\email{\small{sbhati@mit.edu}}

\begin{document}

\maketitle

\section{Abstract}

Large Audio Language Models (LALMs) deliver strong performance across speech and audio tasks, but their audio encoders generate high-rate token sequences (e.g., 25 tokens/s), making attention computation costly and limiting scalability. In this paper, we explore techniques such as unsupervised segmentation, uniform average pooling, etc., to reduce the number of audio tokens before they are consumed by the LLM decoder. To mitigate potential performance degradation, we employ low-rank adapters during finetuning. We evaluate our proposed models on two tasks, automatic speech recognition and speech-to-speech translation tasks, that are dependent on effectively uncovering  the underlying lexical content of the input signal, and study the effect of downsampling on these tasks. Experimental results show that compressed LALMs can achieve performance closer to frame-level LALMs while reducing the input audio token count up to three times before the LLM backbone.

\section{Introduction}

Audio is a primary medium of human communication and interaction, and a rich understanding of audio signals is essential for building systems that can engage naturally with humans and operate effectively in real-world environments. Large Language Models (LLMs)~\cite{devlin2018bert,raffel2020exploring,brown2020language,ouyang2022training,zhang2022opt,zhao2023survey} extended with audio inputs, commonly referred to as Large Audio Language Models (LALMs), have recently achieved remarkable success in audio understanding and reasoning~\cite{deshmukh2023pengi,gong2023listen,gong2023joint,tang2023salmonn,ghosh2024gama,chu2024qwen2,huang2024audiogpt,ding2025kimi,ghosh2025audio,goel2025audio, rouditchenko2025omni}.

\begin{figure}
    \centering
    \includegraphics[width=0.9\linewidth]{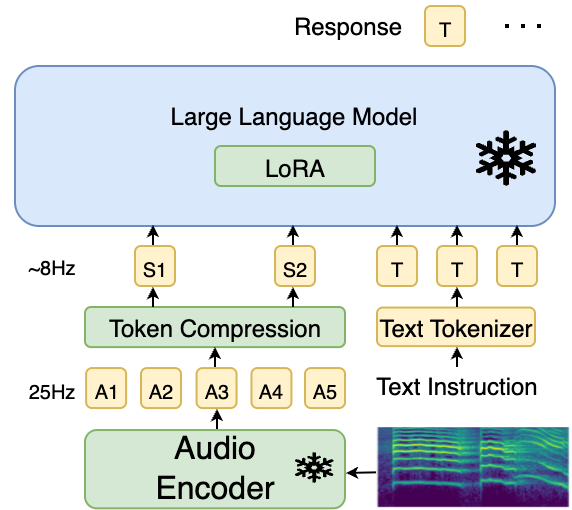}
    \caption{Overview of the audio token compression in large audio language model. The compression module takes the output from the audio encoder and compresses the audio tokens, which are fed into the LLM.}
    \label{fig:tok_comp_overview}
    \vspace{-8pt}
\end{figure}

A typical LALM has the following components: (1) an audio encoder, often a pretrained model such as Whisper, which extracts meaningful features (or tokens) from the audio signal, (2) a language model backbone, which reasons over these embeddings and generates a response, and (3) optionally, a speech synthesizer to generate a spoken response. Advancements in computation and the availability of vast amounts of labeled data have enabled LALMs to achieve strong performance across a wide range of audio tasks. However, the majority of the current LALMs focus on short audio, typically 10 seconds, primarily because the audio encoders produce features at a very high rate. For instance, a 10-second audio clip, the audio encoder can yield around 500 features (50 features/second), resulting in extremely long sequences. Since transformer-based LLMs scale quadratically with sequence length, such long input sequences pose significant computational challenges.

Existing approaches attempt to mitigate this issue by either uniformly pooling the audio features~\cite{chu2024qwen2,ghosh2025audio,xu2025qwen2,rouditchenko2025omni} or transcribing speech into text~\cite{huang2024audiogpt}. Uniform pooling (typically by a factor of two) reduces token counts, but the number of audio tokens remains significantly higher than that of text tokens. A significant token reduction can be achieved by using the text transcription instead of the audio embeddings. Transcription, however, discards important paralinguistic information such as speaker identity, prosody, emotion, and environmental context. Here, we explore unsupervised unit discovery, uniform average pooling, and uniform downsampling to compress audio tokens and reduce the number of audio tokens being sent to the LLM backbone. We explore how these methods affect the recovery of the underlying lexical content. 

Unsupervised unit discovery aims to segment audio into acoustically homogeneous units such as phone or word-like segments~\cite{bhati2021segmental, cuervo2022contrastive,cuervo2022variable,van2020vector,kamper2017embedded}. Segment boundaries can be used to merge frame-level features to generate segment-level features. Prior work has shown that segment-level features can achieve performance comparable to frame-level features in phoneme classification, while requiring significantly fewer features(tokens). Frameworks such as Segmental Contrastive Predictive Coding~\cite{bhati2021segmental} and variable-rate Contrastive Predictive Coding~\cite{cuervo2022variable} demonstrate the effectiveness of unsupervised boundary detection in producing multiscale representations at both the frame and phone level. We use unsupervised unit discovery to discover segments and generate segmental audio features. These segmental features preserve the underlying lexical while reducing the number of audio tokens before the LLM.

While using the compressed audio tokens reduces the audio token count substantially, the shift from frame-level to compressed representations disrupts alignment between the audio embedding space and the language model input space. To address this, we add low-rank adapters to the LLM and finetune to restore alignment. Our experiments demonstrate that only a small amount is sufficient to recover most of the performance of frame-level LALMs. We benchmark our models on speech recognition and speech-to-speech translation tasks using CommonVoice~\cite{ardilacommon} 4 and CoVoST~\cite{wang2020covost} 2 datasets and show that we can achieve good performance while reducing the number of tokens significantly. We chose these tasks as they depend on discovering the underlying temporal lexical content accurately. Some other tasks such as environment or speaker classification might allow even more aggressive down-sampling.  

\section{Related Work}

LLMs trained on large-scale text data achieve remarkable performance on a wide variety of reasoning and understanding tasks. LLMs learn general-purpose representations that can be aligned to a desired task via instruction tuning~\cite{zhang2023instruction}. Large Audio Language Models(LALMs) extend LLMs beyond text to reason and understand speech and general audio.

Pengi~\cite{deshmukh2023pengi} used a hierarchical transformer-based HTSAT~\cite{chen2022hts} as the audio encoder, CLIP text encoder, and GPT2~\cite{radford2019language} as the LLM backbone. SALMONN~\cite{tang2023salmonn} uses dual encoders: a Whisper encoder to extract speech information and a BEATs encoder~\cite{chen2022beats} to extract non-speech audio semantics information. The embeddings from the two encoders are concatenated before the LLaMA LLM backbone. LTU~\cite{gong2023listen} uses AST~\cite{gong2021ast} as the audio encoder and the LLaMA as the language model. LTU shows free-form open-ended question-answering capabilities and outperforms the existing LALMs on the closed-ended tasks. LTU-AS extended the LTU framework beyond audio to jointly understand spoken text and non-speech audio. LTU-AS used a Whisper encoder to extract audio representations and the transcripts generated by the Whisper model as input to the LLaMA backbone. AudioGPT~\cite{huang2024audiogpt} extended ChatGPT beyond text to handle complex audio and speech tasks by using existing speech and audio models. AudioGPT analyzes the user prompt and assigns a model based on the prompt. For example, for the speech recognition task, whisper~\cite{radford2023robust} is used, whereas for sound detection, a Pyramid Transformer~\cite{xin2022audio} is used.

GAMA~\cite{ghosh2024gama} builds upon LTU and improves the audio embeddings. GAMA combines the information extracted from various layers from AST via an Audio Q-former instead of just using the output from the last AST layer. GAMA also proposed and used an improved instruction tuning dataset to improve complex reasoning. Qwen2-Audio~\cite{chu2024qwen2} used a whisper encoder~\cite{radford2023robust} for encoding speech and audio data, and the Qwen2 LLM backbone. Qwen2-Audio used massive amounts of speech, audio, and music data and fully finetuned both the audio encoder and LLM backbone. Qwen2-Audio achieved state-of-the-art performance on a wide variety of audio and speech tasks, such as speech recognition, speech translation, and audio and music understanding. Audio Flamingo (AF)~\cite{kong2024audio,ghosh2025audio,goel2025audio} models focused on improving the audio encoder representations and improving the complex audio reasoning and understanding. AF2~\cite{ghosh2025audio} trained an improved CLAP audio encoder to improve long audio understanding. AF3 trained an AF-Whisper to improve Whisper performance on audio and music-related tasks. AF3~\cite{goel2025audio} achieves state-of-the-art performance on a wide variety of audio and speech benchmarks.

\begin{table*}
    \centering
    \begin{tabular}{ccccccccccc|ccccc} \toprule
         \multicolumn{2}{r}{tok/s (rel. flops) }&\multicolumn{9}{c|}{ASR (WER $\downarrow$)} & \multicolumn{5}{c}{S2TT (BLEU $\uparrow$)} \\ \midrule
         & &\multicolumn{2}{c}{LibriSpeech} & \multicolumn{6}{c}{CommonVoice} & Avg & \multicolumn{2}{c}{X $\rightarrow$ en} & \multicolumn{2}{c}{en$\rightarrow$X} & Avg\\ \cmidrule{2-16}
         & & t-c & t-o & de & en & es & fr & it & zh & & de & zh & de & zh & \\ \midrule
         Q2A & 25 (1) & 1.7 & 4.0 & 7.7 & 7.1 & 5.4 & 8.3& 6.6& 6.5  & 5.9 &33.5 & 23.9 & 29.6 & 45.5 & 33.1\\ 
         +LoRA & 25 (1) & 1.7 & 3.9 & 7.7 & 7.0 & 5.4 & 8.2 & 6.6 & 5.5 & 5.8 & 33.9 & 24.3 & 30.2 & 46.3 & 33.7\\ \midrule
         
        UniAvg$*$  & 12.5 (0.62) & 8.0 & 11.2 & 17.1 & 14.9 & 13.9 & 16.8 & 16.2 & 15.2 & 14.2 & 23.3 & 14.8 & 21.3 & 33.7 & 23.3\\
        UniAvg$*$ & 8.33 (0.5) & 29.8 & 33.6 & 45.4 & 42.9 & 41.9 & 44.4 & 45.8 & 42.6 & 40.8 & 11.1 & 6.9 & 8.6 & 16.0 & 10.7\\ \midrule
         
         UniAvg & 12.5 (0.62) & 1.9 & 4.3 & 8.8 & 7.7 & 6.2 & 9.2 &7.6 & 6.2 & 6.5 & 32.1 & 22.6 & 29.0 & 44.4 & 32.0\\
         UniSamp & 12.5 (0.62) & 2.4 & 5.5 & 9.6 & 8.2 & 6.7 & 10.0 & 8.4 & 6.7 & 7.2 & 32.0 & 22.9 & 28.8 & 44.3 & 32.0\\ \midrule
         UniAvg & 8.33 (0.5) & 2.2 & 4.9 & 9.7 & 8.2 & 6.9 & 10.1 & 8.6 & 6.8 & 7.2 & 31.1 & 21.5 & 28.3 & 43.2 & 31.0\\
         UniSamp & 8.33 (0.5) & 4.6 & 8.5 & 12.2 & 9.6 & 8.6 & 12.8 & 11.0 & 9.1 & 9.6 & 29.1 & 19.6 & 27.0 & 41.9 & 29.4\\
         UnSeg & 8.23 (0.51) & 2.7 & 5.2 & 9.7 & 8.3 & 6.7 & 9.8 & 8.5 & 6.6 & 7.2 & 31.9 & 22.7 & 28.7 & 43.8 & 31.8\\
         \bottomrule
    \end{tabular}
    \caption{ASR and S2TT performance for different compression modules added to Qwen2Audio (Q2A) LALM. t-c: test-clean, t-o: test-other, UniAvg: Uniform average pooling, UniSamp: Uniform sampling, UnSeg: Unsupervised Segmentation, $*$: No LoRA/no training}
    \label{tab:main_res}
    \vspace{-8pt}
\end{table*}

\section{Audio token compression in large Audio Large Language Models}

\textbf{Audio Encoder:} We use the audio encoder from the Qwen2-Audio LALM~\cite{xu2025qwen2}, which is based on the Whisper-large v3 model~\cite{radford2023robust}. 
To preprocess the audio, we resample the audio to 16KHz and extract a 128-dimensional mel-spectrogram using a 25ms window and a 10ms stride. Each feature in the spectrogram corresponds to a 10ms segment of the audio signal, or approximately 100 features per second. In the end, a pooling layer with a pooling rate of two is used to reduce the number of audio features. The final features cover approximately 40ms of the audio. The audio encoder generates 50 features per second.
 
\textbf{Unsupervised Segmentation:}
We use a simple boundary detector proposed in Segmental CPC~\cite{bhati2021segmental}. The frames that belong to the same underlying unit have higher similarity to each other, and if adjacent frames have low similarity, then it might be an indicator of a segment change. $\mathbf{d}=(d_1,d_2,...,d_{L-1})$ captures the dissimilarity between two adjacent frames,
 $d_{t} = 1 - \mathrm{sim}(\zvi{t},\zvi{t+1})$ where $\zvi{t}$ is the output of the audio encoder at t$^{th}$ time step and $L$ is the total number of features. To locate the segment boundaries, we find the frame indexes with high similarities to their neighbors. To achieve this, a simple peak detector is used. We define $p_t$ as a peak if it is higher than both its left and right neighbours, i.e., $d_t > d_{t-1}$ and $d_t > d_{t+1}$.

\textbf{Local pooling:} We uniformly downsample the audio features by a factor of $K$ and generate $25/K$ features per second. We use the following two methods: 1) \textbf{Uniform average pooling:} by using an average pool kernel with $K$ kernel size and $K$ stride, 2) \textbf{Uniform Sampling:} By uniformly sampling every $K^{th}$ feature frame from the audio features.

\textbf{Low-Rank Adapters:} To re-align the modified audio embeddings and the LLM embeddings space, we use Low Rank Adapters (LoRA)~\cite{hu2021lora} to finetune the LLM instead of finetuning the entire LLM. LoRA adds a small set of learnable weights that can be decomposed into a product of two low-rank matrices on top of the pre-trained LLM weights. The final LLM parameters are the addition of the learned low-rank matrices and frozen LLM parameters. This allows us to modify the LLM without learning all the parameters directly. For Qwen2 LLM, we add LoRA adapters with rank 16 and $\alpha=$32 to the key and query projection layers in all the attention layers. We do not add any adapters to the audio encoder. This adds approximately 9M learnable parameters to the model.

\section{Experiments}

We use the Qwen2-Audio-7B~\cite{chu2024qwen2} model and add token compression modules to it. We experiment with uniform average pooling, uniform sampling, and unsupervised segmentation as the compression modules. 


We use LibriSpeech~\cite{panayotov2015librispeech} and CommonVoice~\cite{ardilacommon} version 4 datasets for generating the training and evaluation data for the Automatic Speech Recognition (ASR) task. We use English (en), Spanish (es), French (fr), Italian (it), German (de), and Mandarin (zh) from the CommonVoice dataset. We use the Covost2 dataset~\cite{wang2020covost}, specifically de, en, and zh, and en translation pairs for generating training and evaluation data for the speech-to-text translation (S2TT) task.

We train the models for approximately 5000 steps with a batch size of 5 per GPU. We also accumulate the gradients for 2 steps, and thus the effective batch size is 40. We use a learning rate of 2e-4 and use 2000 warm-up steps. We use 4 A6000 GPUs for all our experiments. We train 0.1$\%$ of the model parameters via LoRA. 
During training, the output label contains the language identifier followed by the ASR or S2TT output. During the inference, the question contains the language id, and the model is expected to return only the ASR or the S2TT outputs.  

\begin{figure}
     \centering
     \begin{subfigure}[b]{0.24\textwidth}
         \centering
         \includegraphics[width=\textwidth]{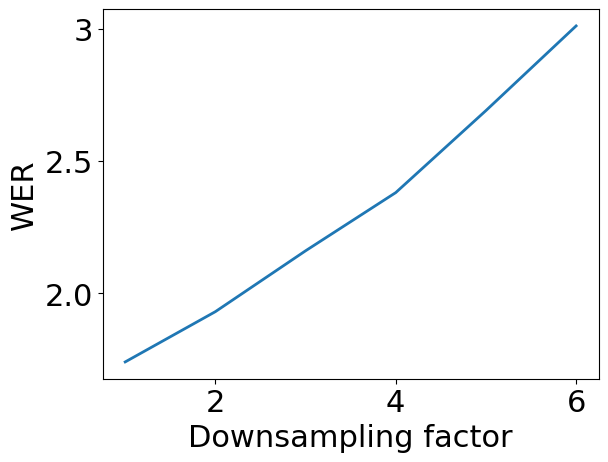}
         \caption{UniAvg}
         \label{fig:y equals x}
     \end{subfigure}%
     ~
     \begin{subfigure}[b]{0.24\textwidth}
         \centering
         \includegraphics[width=\textwidth]{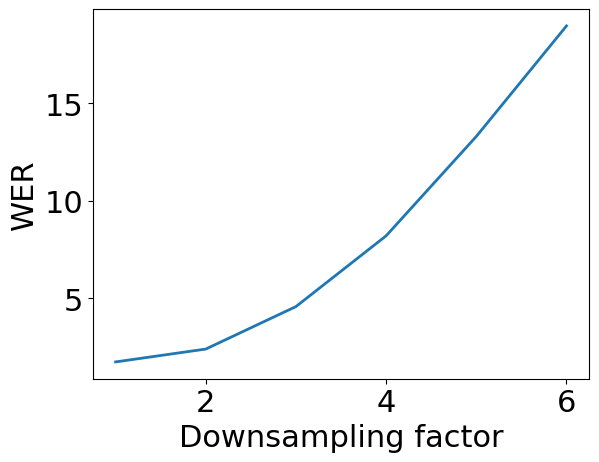}
         \caption{UniSamp}
         \label{fig:three sin x}
     \end{subfigure}
        \caption{WER for UniAvg and UniSamp vs compression factors}
        \label{fig:WERvsCompFactor}
        \vspace{-10pt}
\end{figure}

\subsection{Performance on ASR and S2TT task}
We use the Qwen2-Audio model as the baseline, and as seen in Table~\ref{tab:main_res}, the Qwen2-Audio model achieves strong performance on both ASR and S2TT across all languages. We also train an improved Qwen2-Audio baseline where we add LoRA and fine-tune the model on the training data. We observe minimal improvement over the baseline performance. It is possible that the Commonvoice and the Covost data are already included in the training data for Qwen2-Audio, and further finetuning on these datasets does not improve performance.

We then explore the impact of compressed tokens on model performance without adding any LoRA modules or any finetuning. As shown in Table~\ref{tab:main_res}, for both compression factor of 2 and 3, both the ASR and S2TT performance degrades. For larger compression factors i.e. 3, the degradation is severe, WER $~$40.  and it renders the model useless. This shows the importance of LoRA modules and realigning the compressed audio embeddings and the LLM. As we discuss below, LoRA allows us to recover most of the original model performance. Token compression also allows us to reduce computation cost (relative flop count) of the model to half, achievable with a compression factor of 3, without major performance loss which is significant for model training and deployment.  

As shown in Table~\ref{tab:main_res}, we can significantly reduce the token rate while achieving strong performance on both tasks across all languages. For a small increase in WER i.e., $<1$, we can reduce the number of audio tokens in half. Reducing the number of tokens to a third further increases the WER, but the WER increase is still pretty small, $<1.5$. The uniform average pooling outperforms the uniform sampling in terms of both WER and BLEU scores. 
Unsupervised segmentation outperforms uniform sampling and performs slightly better than uniform average pooling. Based on the results, token merging, either based on unsupervised segmentation or uniform average, outperforms token sampling (or token dropping). We hypothesize that averaging keeps the information, whereas sampling loses some information.

Next, we explore how far we can push the compression rates. We vary the compression rates for uniform sampling and uniform averaging. Figure~\ref{fig:WERvsCompFactor} shows the impact of compression rate on the ASR performance on the Librispeech dataset. For both the uniform average pooling and uniform sampling, the error increases as the compression factor increases. The WER increases significantly for the uniform sampling compared to the uniform sampling, which further strengthens our hypothesis that audio token merging is better than token sampling for compressing audio tokens. For a uniform average, the error rate doubles while we can reduce the number of audio tokens by a factor of six. The WER, even at a very high compression rate, is still reasonably low. We can select a desired compression factor depending on the trade-off between the available computational resources and WER.

\subsection{Does realignment on one language transfer to other languages}

\begin{table*}[h!]
\centering
\begin{tabular}{ccccccccccccccc} \toprule
 & \multicolumn{9}{c}{ASR (WER $\downarrow$)} & \multicolumn{5}{c}{S2TT (BLEU $\uparrow$)} \\ \midrule
 & \multicolumn{2}{c}{LibriSpeech} & \multicolumn{6}{c}{CommonVoice} & \multicolumn{1}{l}{Avg} & \multicolumn{2}{c}{X $\rightarrow$ en} & \multicolumn{2}{c}{en$\rightarrow$X} & \multicolumn{1}{l}{Avg} \\ \cmidrule{2-15}
 & t-c & t-o & de & en & es & fr & it & zh & \multicolumn{1}{l}{} & de & zh & de & zh & \multicolumn{1}{l}{} \\ \midrule
Librispeech & 2 & 4.4 & 14.0 & 12.1 & 10.9 & 14 & 13 & 8.8 & 9.9 & 24.9 & 14.8 & 22 & 34.4 & 24.0 \\
+CV\_en & 2.0 & 4.4 & 9.9 & 8.1 & 7.1 & 10.6 & 8.9 & 6.8 & 7.2 & 25.0 & 14.8 & 24.0 & 38.1 & 25.5 \\
+CV\_all & 2.1 & 4.6 & 9.3 & 8.1 & 6.6 & 9.8 & 8.1 & 6.5 & 6.9 & 23.9 & 14.5 & 23.8 & 37.9 & 25.0 \\
+covost & 2.1 & 4.8 & 9.7 & 8.2 & 6.8 & 10.1 & 8.5 & 6.7 & 7.1 & 31.1 & 21.5 & 28.3 & 43.2 & 31.0 \\ \bottomrule
\end{tabular}
\caption{Impact of training datasets on the performance with uniform average pooling with compression factor$=$3}
\label{tab:data_exp}
\vspace{-6pt}
\end{table*}

The Qwen2-Audio model supports multiple languages and tasks. Here, we explore whether realigning in one language improves performance in other languages. We begin by just using LibriSpeech (English) for training and evaluating performance across other languages. As shown in Table~\ref{tab:data_exp}, using only the LibriSpeech dataset, the models perform reasonably well across different languages. Next, we add English(en) from CommonVoice, which further reduces the WER across all languages. We hypothesize that training on English from CommonVoice datasets enables the model to learn the CommonVoice distribution and perform better across all languages. Adding training data from the other languages further reduces the WER on all the languages and achieves the best performance.

\subsection{Does realignment on one task transfer to other tasks}

Previous experiments showed that training on a task, ASR, for example, transfers across languages and improves performance on unseen languages. In this section, we explore whether realignment on one task, i.e., ASR, improves performance on other tasks such as S2TT.

We begin by training the model on the ASR task on the Librispeech dataset. The model performs reasonably well on the speech-to-speech translation task, Table~\ref{tab:data_exp}. However, unlike the ASR task, adding CommonVoice data, either from English or other languages, does not improve performance on the S2TT task. Adding COVOST data or training the model on the S2TT task yields the best performance.

\subsection{Learnable compression modules}
Instead of using average pooling as the compression module, we explore if using a learnable layer leads to better performance. We use a convolution layer for compressing the audio tokens. This significantly increases the overall number of trainable parameters (0.14$\%$ and 0.16$\%$ for compression factor 2 and 3 respectively) but increases the learning capacity of the compression module. As shown in Table~\ref{tab:learnable_downsamp}, using a learnable layer leads to worse performance. It is possible that the trainable compression modules require more training which we plan to explore in future.

Currently, the compression module is added after the audio encoder has processed at audio and before the LLM backbone. We add the compression module before the audio encoder which allows the audio encoder to operate at compress tokens. This can further reduce the overall computation used by the model. However, as seen in Table~\ref{tab:learnable_downsamp} the leads to higher WER.

\begin{table}[h!]
    \centering
    \begin{tabular}{lccc} \toprule
           & tok/s & t-c & t-o \\ \midrule
        UniAvg & 12.5 & 1.9 & 4.3 \\
        Conv & 12.5 & 2.3 & 5.0 \\
        UniAvg & 8.33 & 2.2 & 4.9 \\
        Conv & 8.33 & 2.4 & 5.2 \\
        \midrule
        Before audio encoder & 12.5 & 2.5 & 6.6 \\
        Before audio encoder & 8.33 & 9.3 & 23.4 \\ \bottomrule
    \end{tabular}
    \caption{Using a learnable layer. conv: convolution layer.}
    \label{tab:learnable_downsamp}
\end{table}

\subsection{Impact of WER on understanding and meaning}
\begin{table}[]
    \centering
    \begin{tabular}{lccccc} \toprule
         & tok/s & WER & meaning & readability & mpn \\ \midrule
         Q2A & 25 & 5.9 & 4.89 & 4.97 & 4.93 \\
         UniSamp & 12.5 & 7.2 & 4.79 & 4.94 & 4.90 \\
         UniAvg & 12.5 & 6.5 &4.88 & 4.95 & 4.91 \\
         UniSamp & 8.33 & 9.6 &4.44 & 4.90 & 4.79 \\
         UniAvg & 8.33 & 7.2 & 4.84 & 4.93 & 4.89 \\
         UnSeg & 8.23 & 7.2  & 4.83 & 4.93 & 4.89 \\ \bottomrule
    \end{tabular}
    \caption{Average scores across meaning, readability, and misspelled proper nouns (mpn)}
    \label{tab:gptevalscores}
    \vspace{-10pt}
\end{table}
LALMs are capable of understanding and reasoning over the input audio. For many reasoning tasks, the exact transcription does not matter as long as the meaning and the readability of the text are preserved. WER penalizes all errors in the prediction equally, regardless of their impact on the readability, meaning. Some other works have also focused on humanizing the word error rate and have focused on evaluating the readability and meaning~\cite{kim2021evaluating,hwer}. Instead of using human annotators, we use ChatGPT to evaluate the predictions from the LALM and the reference along the following three metrics:

\begin{itemize}
    \item meaning: Does the prediction change the meaning compared to the reference?
    
    \item readability: Does the prediction affect readability compared to the reference?

    \item misspelled proper nouns (mpn): Does the prediction misspell proper nouns that appear in the reference?
\end{itemize}

ChatGPT takes both prediction and reference as input and generates a score between 1 and 5 for each of the three metrics. As seen from the Table~\ref{tab:gptevalscores}, for uniform average with compression factor 2 or 3 and for unsupervised segmentation, even though the WER increases, the meaning and readability do not decrease significantly. Only when the WER increases significantly, for uniform sampling, does the meaning change, and scores decrease significantly.

\section{Conclusions and Future Work}

In this paper, we explore token compression in large audio language models to reduce the number of audio tokens generated by the audio encoder. We use LoRA to align the compressed audio tokens to the LLM embedding space. We can reduce the number of audio tokens by up to 3 times without significantly worsening the performance on the ASR and S2TT tasks. 

In the future, we would like to explore methods that do not require training the model with additional LoRA weights for rebalancing the attention weights. We would also like to benchmark the compressed LALMs on other tasks, such as audio classification and understanding. 

\section{Generative AI Use Disclosure}
Generative AI was only used for basic editing such as grammar checks and improvements. 

\bibliographystyle{IEEEbib}
\bibliography{strings}

\end{document}